\documentclass[conference]{IEEEtran}
\IEEEoverridecommandlockouts
\usepackage{cite}
\usepackage{amsmath,amssymb,amsfonts}
\usepackage{algorithmic}
\usepackage{graphicx}
\usepackage{textcomp}
\usepackage{xcolor}
\usepackage{bbding}
\usepackage{pifont}
\usepackage[a4paper, total={184mm,239mm}]{geometry}
\def\BibTeX{{\rm B\kern-.05em{\sc i\kern-.025em b}\kern-.08em
    T\kern-.1667em\lower.7ex\hbox{E}\kern-.125emX}}
\begin{document}

\title{AFPR-CIM: An Analog-Domain Floating-Point RRAM-based Compute-In-Memory Architecture with Dynamic Range Adaptive FP-ADC\\}

\author{
    \IEEEauthorblockN{Haobo Liu\IEEEauthorrefmark{1} \IEEEauthorrefmark{2}, Zhengyang Qian\IEEEauthorrefmark{2}, Wei Wu\IEEEauthorrefmark{2}, Hongwei Ren\IEEEauthorrefmark{2
}\IEEEauthorrefmark{3
}, Zhiwei Liu\IEEEauthorrefmark{1} and Leibin Ni\IEEEauthorrefmark{2}
    \thanks{This work was supported in part by the STI 2030-Major Projects (2021ZD0201205) and Nature Science Foundation of China under grant No.
61974017. Corresponding authors: Leibin Ni (nileibin@huawei.com), Zhiwei Liu (ziv\_liu@hotmail.com)}}
    \IEEEauthorblockA{\IEEEauthorrefmark{1} Shenzhen Institute for Advanced Study, University of Electronic Science and Technology of China, Shenzhen, China}
    \IEEEauthorblockA{\IEEEauthorrefmark{2} Central Research Institute, 2012 Laboratories, Huawei Technologies Co., Ltd., Shenzhen, China}
    \IEEEauthorblockA{\IEEEauthorrefmark{3} MICS Thrust, The Hong Kong University of Science and Technology (Guangzhou), Guangzhou, China}
}

\maketitle

\begin{abstract}

Power consumption has become the major concern in neural network accelerators for edge devices. The novel non-volatile-memory (NVM) based computing-in-memory (CIM) architecture has shown great potential for better energy efficiency. However, most of the recent NVM-CIM solutions mainly focus on fixed-point calculation and are not applicable to floating-point (FP) processing. In this paper, we propose an analog-domain floating-point CIM architecture (AFPR-CIM) based on resistive random-access memory (RRAM). A novel adaptive dynamic-range FP-ADC is designed to convert the analog computation results into FP codes. Output current with high dynamic range is converted to a normalized voltage range for readout, to prevent precision loss at low power consumption. Moreover, a novel FP-DAC is also implemented which reconstructs FP digital codes into analog values to perform analog computation. The proposed AFPR-CIM architecture enables neural network acceleration with FP8 (E2M5) activation for better accuracy and energy efficiency. Evaluation results show that AFPR-CIM can achieve 19.89 TFLOPS/W energy efficiency and 1474.56 GOPS throughput. Compared to traditional FP8 accelerator, digital FP-CIM, and analog INT8-CIM, this work achieves 4.135×, 5.376×, and 2.841× energy efficiency enhancement, respectively.

\end{abstract}


\section{Introduction}
Next-generation information technologies like edge computing and cloud computing require large amounts of data processing and data transmission and have an increasing demand for computation capacity and energy efficiency for hardware, especially AI accelerators.

As the AI tasks get more complex, the floating-point (FP) format becomes more and more popular in the training and inference phases. In comparison to the INT format with the same bit width, the FP format has a larger dynamic range with a higher hardware cost \cite{ren2023ttpoint}. Currently, a number of well-known accelerators, like Google TPU and NVIDIA GPU, have created deep optimization implementations of the BF16 format\cite{wang2019bfloat16}. The bit reduction of the FP format to 8-bit has led to the emergence of the low-bit FP format as a promising alternative for DNN quantization\cite{park20219,van2023fp8}. NVIDIA has announced the adoption of the FP8 format for the transformer engine of its new Hopper Architecture GPUs. FP8, an extension of FP16, minimizes model size and inference cost. Due to non-linear sampling of real numbers, FP8 outperforms INT8 in inference. However, several prior studies on the FP8 format have primarily focused on the algorithmic level\cite{kuzmin2022fp8}. The bottleneck limiting the development of low-precision FP formats lies in the optimization of software algorithms on the one hand, but mainly in the power consumption overhead at the FP8 hardware level. FP8 (E2 to E5) is compared to INT8 in terms of algorithms and hardware \cite{van2023fp8,kuzmin2022fp8}. The FP8 format has better software efficiency, while it is accompanied by much higher hardware power consumption compared to INT8. An in-depth study and analysis of FP hardware is imperative. However, most FP8 accelerators are still based on traditional Von Neumann architecture with digital computation, resulting in limited improvement in energy efficiency\cite{park20219,van2023fp8,kuzmin2022fp8}.

In recent years, NVM-based analog CIM, such as resistive random-access memory (RRAM), has gained recognition for its notable energy efficiency \cite{shafiee2016isaac,chi2016prime,yu2018neuro}. Analog Multiply-Accumulate (MAC) operation is directly performed in the memory array, avoiding numerous data movements to achieve high energy efficiency. Meanwhile, binary RRAM-based CIM designs are unable to take advantage of the multi-bit properties of multi-level-cell (MLC) devices\cite{ni2016energy,ni2017energy}. However, the majority of current NVM-CIM solutions primarily concentrate on fixed-point computation and are not applicable to FP processing. This is mainly limited by the fact that discrete FP numbers and their separated bit parts are difficult to match with the continuous process in analog computation \cite{shafiee2016isaac,chi2016prime,yu2018neuro,ni2016energy,ni2017energy,wan2022compute,liu202033,zhang2020robust}.

In this work, we propose the AFPR-CIM system, which presents a new architecture based on the physical paradigm of the conventional analog CIM. We demonstrate that the proposed architecture is more appropriate for the initial goal of the bit reduction from FP16 to FP8 format. It may clearly highlight the parallelism and outstanding energy efficiency of analog computing. The key contributions of this work are as follows:
\begin{itemize}
\item We propose an all-analog domain CIM architecture for floating-point (FP8) format calculations based on RRAM devices to achieve better energy efficiency and neural network accuracy. We also analyze the FP8 bit assignments (E2M5, E3M4) versus the INT8 format in terms of hardware efficiency and network accuracy, and determine the optimal design for the AFPR-CIM system. An energy efficiency of 19.89 TFLOPS/W and a throughput of 1474.56 GFLOPS are achieved.
\item We propose an adaptive dynamic range FP-ADC that achieves adaptive matching of the input dynamic range through automatic capacitive reconfiguration and charge sharing. We enable the ADC to naturally convert the analog domain MAC results into FP (E2M5) digital codes through the capacitor combination. 
\item We propose an FP-DAC that reconstructs the FP activation codes into analog input values to perform analog domain MAC. The programmable analog gain is utilized to provide an analog representation of the FP's exponent without additional overhead.
\end{itemize}

\section{Related Works}

In general, FP hardware research based on CIM outperforms the Von Neumann architecture in energy efficiency and area. Currently, most FP-CIM works are implemented in the digital domain with FP32 or FP16 format\cite{tu202228nm}. The primary method of the digital approach is using a significant number of digital modules to perform FP calculations. There are also some works that use RRAM to form logic gate circuits for FP computations\cite{lu2021rime}. The above implementations in digital domains are limited in their computational parallelism due to routing congestion. And the exponential bit inevitably leads to power consumption due to alignment operations. Additionally, most all-digital FP-CIM solutions are based on volatile memory like SRAM, and cannot be implemented on MLC devices with better area efficiency. 

The analog CIMs are typically realized in crossbars based on physical laws such as Ohm's law and Kirchhoff's current law. The computational results are expressed in analog values like currents. The number of activated wordlines and input patterns may vary in different application scenarios, leading to a wide-distributed dynamic range of MAC results. Traditional readout circuitry needs to cover the whole dynamic range, resulting in overdesign and wasted power consumption. Regarding the issue above, previous research has focused on two primary ideas. One idea is to optimize the computed signals. Work in \cite{chen20237} controls the signal dynamic range to be distributed in the near-zero-mean range by digital processing. Another solution suggested in \cite{wan2022compute} is to use a voltage-domain sensing scheme instead of a current-domain sensing scheme. However, the sampling linearity degrades due to the charge injection. Another idea is to optimize the ADC for the readout stage. \cite{chen20237} designed the ADC in a small range based on the estimated distribution of results. In \cite{liu202033}, ADC is preset to a fixed range by predicting the input signal and the algorithm. However, real-time dynamic range adjustment is still not realized in the above designs.

\section{ AFPR-CIM Architecture with A Novel FP Data Conversion}

AFPR-CIM aims to enable FP8 calculation in the analog domain while preserving the low-power features of analog domain CIM. This will be helpful to alleviate the application bottleneck of high power consumption of FP8 hardware in the digital domain. Unfortunately, due to their discrete features, FP numbers are difficult to compute in the analog domain using effects like Ohm's law. To prevent unavoidable physical mismatches, we proposed the AFPR-CIM architecture based on the traditional analog CIM. It integrates FP-to-INT and INT-to-FP conversion at the Macro interface to achieve the FP-CIM in the analog domain.

\begin{figure}[t]
\centerline{\includegraphics[scale=0.85]{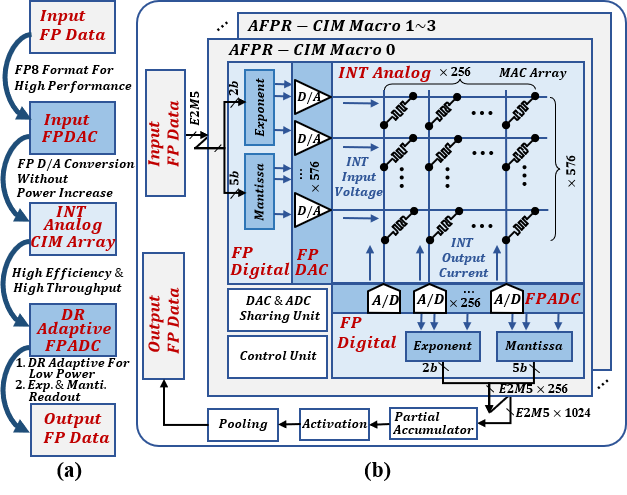}}
\caption{(a) Data flow. (b) The proposed AFPR-CIM architecture.}
\label{fig. 1}
\vspace{-0.5cm}
\end{figure}

\subsection{Analog Floating-Point CIM Architecture}

The overall idea of the whole architecture is to perform an INT physical computation in the analog domain to obtain higher parallelism as well as lower computational energy consumption; At the same time, the neural network is connected via FP numbers in the digital interface between the CIM Macros to represent a larger range and to accommodate the FP format of FP8.

As shown in Fig. \ref{fig. 1}, before inference, the weight data is programmed in the array with multi-level RRAM, represented by device conductance. In the inference phase, the exponent and mantissa bits of the FP8 (E2M5) input activation data are reconstructed into analog input voltages in the INT domain by an FP-DAC, and parallelly input to the RRAM array. According to Ohm's law and Kirchhoff's current law, output currents represent MAC results by input voltages and weight conductances. Then, the analog MAC results in the INT domain are read out through the FP-ADC as the FP digital code consisting of exponent and mantissa bits. This digital code is stored in FP8 format (E2M5). This value could be performed by an activation or pooling operation through an intermediate digital processing unit. In some network mapping cases (to be described later in Network Mapping), the digital processing module will also implement a small portion of summation functions. This architecture realizes the separation of INT and FP domains. Analog computation is performed in the INT domain to take full advantage of its high parallelism and low power consumption, avoiding power wastage due to shift alignment in the all-digital domain. Inputs, outputs, and intermediate processing are performed in the FP domain to take full advantage of its high network efficiency and wide dynamic range.

\begin{figure}[t]
\centerline{\includegraphics[scale=0.72]{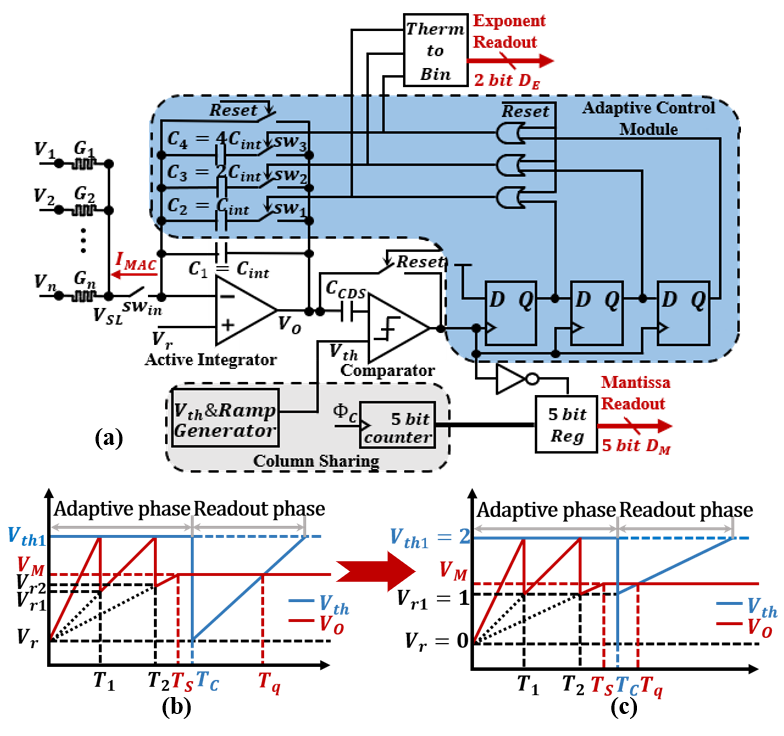}}
\caption{(a) Architecture of dynamic range adaptive FP-ADC. (b) Dynamic range adaptive process. (c) FP conversion process.}
\label{fig. 2}
\vspace{-0.5cm}
\end{figure}

\subsection{Dynamic Range Adaptive FP-ADC}

The most essential part of this work is the dynamic range adaptive floating-point ADC in Fig. \ref{fig. 2}. According to the illustration in Section II, the dynamic range of the MAC result is very different among the multiple SLs. This ADC is designed to accommodate these dynamic ranges without resulting in additional ADC overheads. In the meantime, the physical relationships in this novel ADC enable a natural way to convert fixed-point analog current inputs to FP digital code, thus inspiring the proposed design of the analog domain FP-CIM in this work.

The overall design of the ADC is shown in Fig.  \ref {fig. 2}(a), The resultant current of the array $I_{MAC}$ is integrated into the capacitor $C_{1}$ through an active integrator, and represents the signal as a voltage over the capacitor $C_{1}$. The comparator compares the values of the output voltages $V_O$ and $V_{th}$ to output a pulse. The $V_{th}$ value can be shared by columns. The $C_{CDS}$ are used to compensate for the comparator and integrator offset voltages during reset. The adaptive control module is the most important in ADC. In the adaptive phase, the output of the comparator determines whether the dynamic range needs to be increased, which is realized by setting $V_{th}$. If required, the comparator outputs high, and then adjusts the capacitance combination in the integration circuit, so that it can adaptively follow the resultant current signal. The thermometer code representing the combination of capacitors is converted to a binary code, which is the 2-bit exponent code of the FP readout result. After the capacitor is deployed and the current integration is complete, the result at $V_{O}$ is converted to 5-bit mantissa code in the single slope A/D method. Here the detailed working process will be analyzed:

Fig. \ref{fig. 2}(b) shows the dynamic range adaptive process. Input voltage in each row of the crossbar is represented as $V_{1}$, $V_{2}$ $\cdots$, then the crossbar from the input to the output can be simplified to the structure on the left side of Fig. \ref{fig. 2}(a). RRAM conductance can be expressed as $G_1$, $G_1$ $\cdots$. The positive end of the integral op-amp is connected to the clamp voltage $V_r$. According to the virtual short and Virtual open, $V_{SL}$ is clamped to $V_{r}$. Then the unique resultant current $I_{MAC}$ can be determined, $V_{O}$ is related to $I_{MAC}$ by the equation:

\begin{equation}
I_{MAC}=\sum_{i=1}^{n}\left(V_{r}-V_{i}\right) G_{i}
\end{equation}

In the reset phase, the circuit reset to clear the data and set the voltage at $V_{O}$ to the initial value $V_{r}$. During the adaptive phase, $I_{MAC}$ is integrated into $C_{1}$. When the integrator output voltage $V_{O}$ reaches $V_{th1}$, the high output of the comparator stimulates the first D-flip-flop (DFF) outputs high to control  $sw_{1}$ on, at which time the charge is shared between $C_{1}$ and $C_{2}$ at the moment $T_{1}$. And $V_{O}$ is reduced to $V_{r1}$.

\begin{equation}
V_{r1}=\frac{C_{1}}{C_{1}+C_{2}} V_{t h}+\frac{C_{2}}{C_{1}+C_{2}} V_{r} 
\end{equation}
\begin{equation}
V_{r2}=\frac{C_{1}+C_{2}}{C_{1}+C_{2}+C_{3}} V_{th}+\frac{C_{3}}{C_{1}+C_{2}+C_{3}} V_{r} 
\end{equation}

The current continues to integrate,  similarly, the second DFF output goes high to control $sw_{2}$ on when $V_{O}$ arrives  $V_{th1}$ twice. $V_{O}$ reduces to $V_{r1}$ at the moment $T_{2}$, and so on. At a fixed sample moment $T_{S}$, the value of $V_{O}$ is kept as $V_{M}$. Then perform a single slope A/D conversion. When $V_{th1}$ exceeds $V_{M}$, the comparator outputs high, and the counter number is readout as the mantissa code. The segmentation function of $V_{O}$ with $I_{MAC}$ can be expressed as (4). Therefore, $I_{MAC}$ can be detected indirectly by detecting $V_{O}$.
\begin{equation}
I_{MAC}=\left\{\begin{array}{cc}
\frac{\left(V_{O}-V_{r}\right)}{T}\times C_{1} &{T \in [0,T_1)} 
\\ \frac{\left(V_{O}-V_{r}\right)}{T} \times\left(C_{1}+C_{2}\right) &{T \in [T_1,T_2)}
\\ \frac{\left(V_{O}-V_{r}\right)}{T} \times\left(C_{1}+C_{2}+C_{3}\right)  &{T \in [T_2,T_3)}
\\ \frac{\left(V_{O}-V_{r}\right)}{T} \times\left(C_{1}+ \dots +C_{4}\right) &{T \in [T_3,T_S]}
\end{array} \right. 
\end{equation}
We aim to make this adaptive method able to match the idea of FP expression, which is mainly the nonlinear quantization. The format of the FP number can be expressed as $1.Dmantissa \times 2^{Dexponent}$. $V_{O}$ could vary in a range between 1V and 2V to represent the range of values of $1.Dmantissa$. We set $V_{r}$ to 0 and $V_{th1}$ to 2V, and control all the voltages at the adjustment moments ($V_{r1}$, $V_{r2}$ $\cdots$) drop to 1V. Besides, the integral capacitor combination needs to be set precisely as shown in Fig. \ref{fig. 2}, which is because:

\begin{itemize}
\item According to (4), if $C_1$, $C_2$, $C_3$ and $C_4$ equal to $C_{int}$, $C_{int}$, $2C_{int}$ and $4C_{int}$ respectively with $V_{r}=0$,  the value of $I_{MAC}$ can be derived:

\begin{equation}
I_{MAC}=\left\{\begin{array}{cc}
\frac{C_{int}}{T} \times V_{O} \times 2^0 &{T \in [0,T_1)} 
\\ \frac{C_{int}}{T} \times V_{O} \times 2^1 &{T \in [T_1,T_2)}
\\ \frac{C_{int}}{T} \times V_{O} \times 2^2 &{T \in [T_2,T_3)}
\\ \frac{C_{int}}{T} \times V_{O} \times 2^3 &{T \in [T_3,T_S]}
\end{array} \right. 
\end{equation}

$I_{MAC}$ shows a linear relationship with $V_O \times 2^n$, which gives us great convenience for the expression of floating-point numbers: We just need to correspond $V_{O}$ to $1.Dmantissa$, which is the digital code corresponding to the analog value of the fractional part of $V_{O}$.

\item According to (2) and (3), only this voltage combination is able to realize $V_{r1}=V_{r2}\cdots=1/2(V_{r}+V_{th})=1$.
\end{itemize}

To verify the continuity of the current at the moment of adjustment, we substituted $V_{r1}$ and $V_{th}$ into (4) to compare the current at $T_{1}$. Two currents have the same value, which proves that although the voltage is changing abruptly, the current is still continuous. This is also due to the fact that the total charge of the integral part does not change, but rather is shared between several capacitors. In terms of FP numbers, it is guaranteed that $2\times2^0$ can continuously change to $1\times2^1$ at the edge of adjustment.

When the input current is very small, $V_{O}$ is still not integrated to 1V at  $T_{S}$, the result is not read out. Overall, this novel method effectively realizes dynamic range adaptation and INT-to-FP conversion.

\subsection{Input FP-DAC}

\begin{figure}[t]
\centerline{\includegraphics[scale=0.7]{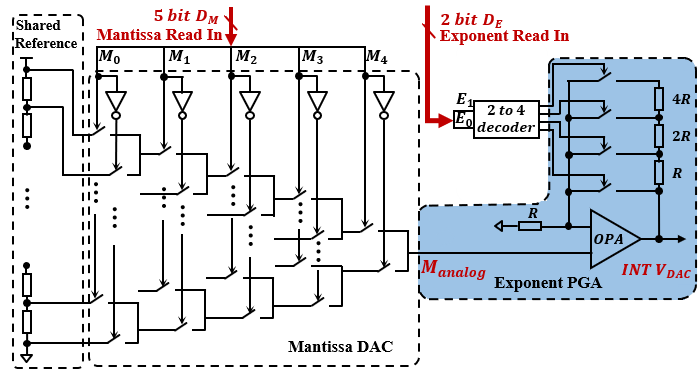}}
\caption{Architecture of Input FP-DAC.}
\label{fig. 3}
\vspace{-0.5cm}
\end{figure}

To adapt the separated design of the INT and FP domain, we designed an FP-DAC that can reconstruct the FP digital activation data in INT analog values before inputting it into the crossbar, as shown in Fig. \ref{fig. 3}. The FP-DAC circuit consists of three parts. The reference module provides a 5-bit reference voltage for the DAC through a resistor network, which can be shared by multiple rows in the array to save power and area. The mantissa DAC is a switching network controlled by the 5-bit mantissa data from the activate. The magnitude of the reference voltage accessed to the back-end PGA is controlled by a combination of switches, which is the analog value Vmantissa corresponding to the mantissa code. The PGA is controlled by the exponential code of the input data and provides programmable gain to the output of the DAC. The exponent of the activation input is converted by a 2-4 decoder into a single control signal, which is used to apply a linear gain of $2^E$ on the resistive PGA by controlling the switches. The closed loop system ensures better linearity of the circuit. As the DAC in traditional analog storage and calculation design also inevitably needs TIA as the input buffer, the PGA in this design also only adds resistors based on the buffer, and at the same time, it realizes the floating-point conversion expression effectively. The output can be expressed as:
\begin{equation}
V_{D A C}=2^{E} \times M_{\text {analog }}
\end{equation}

\subsection{Network Mapping}
\begin{figure}[b]
\centerline{\includegraphics[scale=0.5]{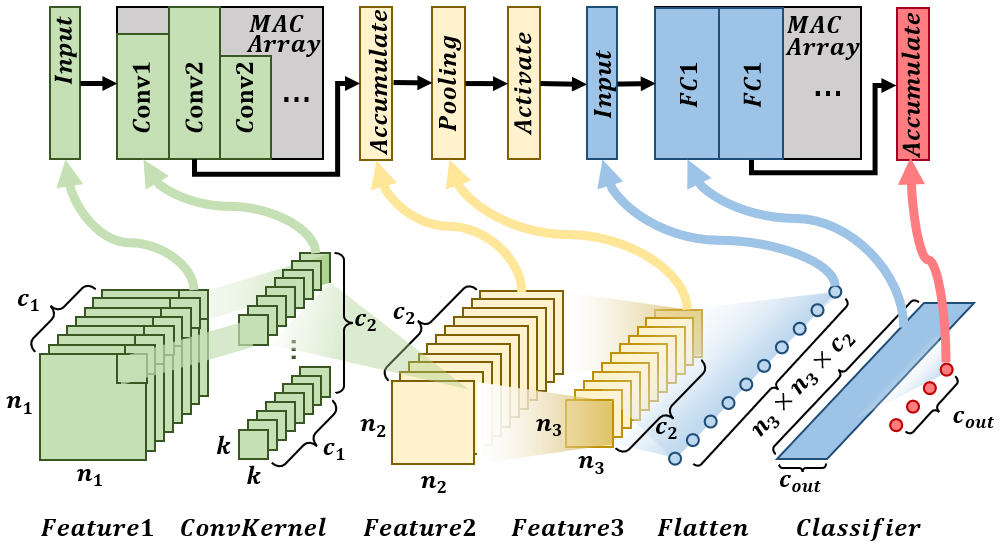}}
\caption{Mapping method for the fully connected layer and the convolution layer on the proposed CIM Macros.}
\label{fig. 4}
\end{figure}

The weight matrix and layer inputs of the convolutional and fully connected layers are mapped to the CIM Macro in the manner of Fig. \ref{fig. 4}. In an NN model, the pooling result of the previous level is used as the input of the next level, and the weights of the convolutional kernel with $C_1 \times k \times k \times C_2$ are converted into a 2D matrix $(C_1\times k\times k) \times C_2$ which is mapped into a crossbar to perform a MAC operation with the inputs of $C_1 \times k \times k$. Similarly, FC weights with a size of $(C_2\times n_3\times n_3)\times C_{out}$ are deployed in the same way for a fully connected layer. When the weight matrix exceeds 576, the result of the MAC operation in the CIM column is a partial sum. We utilize the inter-core routing adder to perform the summation of the partial.

 
\section{Performance Evaluation and Analysis}
The proposed AFPR-CIM scheme is evaluated at the circuit-, Macro- and network- level to demonstrate its functionality and performance. The modeling of RRAM is implemented in the Verilog-A language. The analog and digital supply of the mixed-signal circuit is set to 2.5V and 1.2V, respectively, to cover the 2V floating-point range while minimizing power consumption. The CIM Macro is composed of $576\times 256$ (144K) RRAM devices. We extract the sparsity of the weights from the network model and deploy them into RRAMs using the mapping method described in the previous section. 

\subsection{Functional Analysis}
\begin{figure}[t]
\centerline{\includegraphics[scale=0.28]{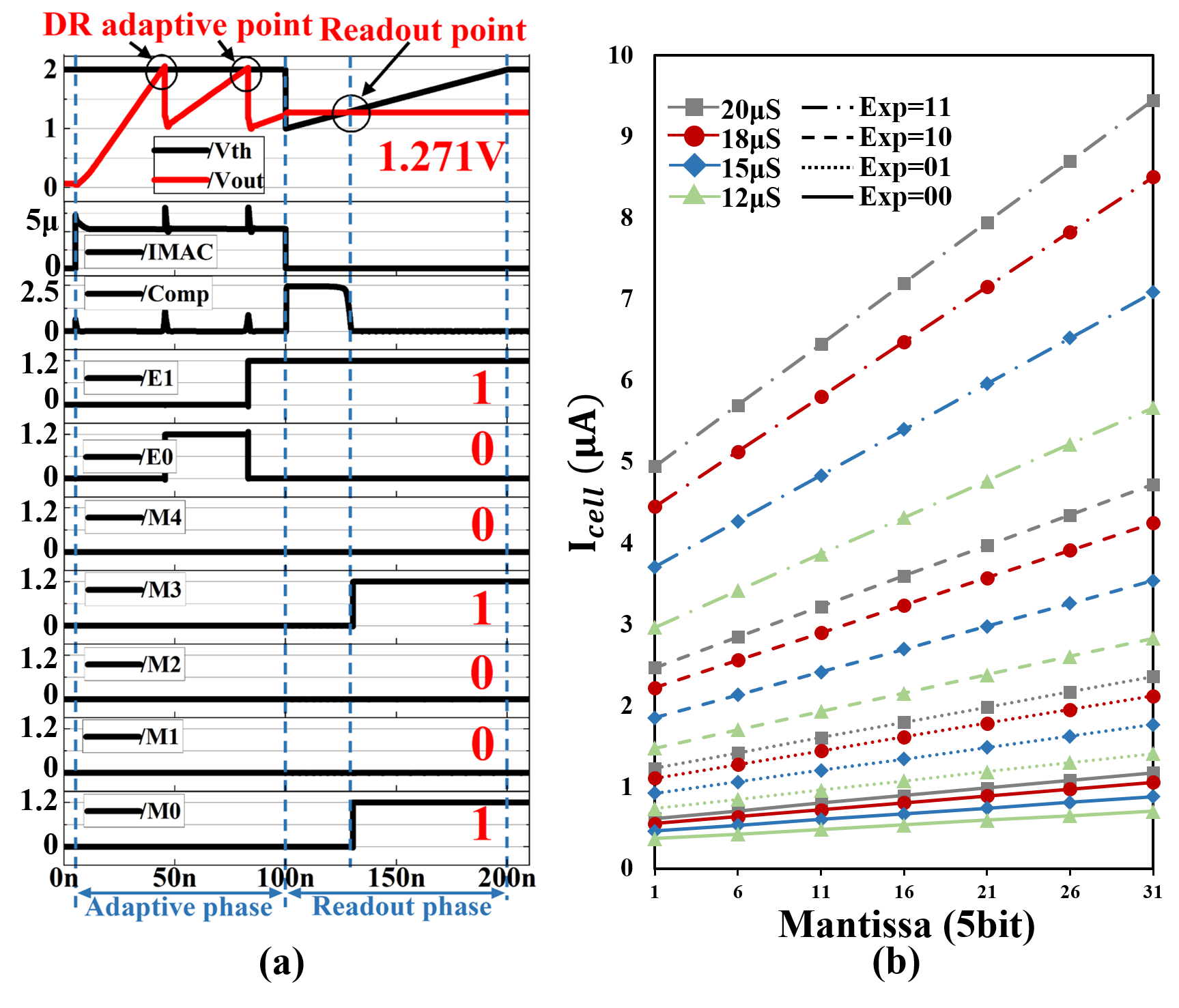}}
\caption{(a) Transient simulation results of FP-ADC. (b) Linearity analysis of FP-DAC.}
\label{fig. 5}
\vspace{-0.5cm}
\end{figure}

We conducted transient simulations to verify the correctness of the proposed idea. The random digital input 1011110 is deployed into the FP-DAC. The analog values converted by FP-DAC input the crossbar where it is multiplied by RRAM conductance with random weights. Fig. \ref{fig. 5}(a) shows the $V_O$, $V_{th}$ transient waveforms. After reset, the integration phase starts at 5ns and remains constant at 5.38uA at each exponent level (shown as the capacitor combination). As shown in the figure, the system adaptively adjusts the dynamic range twice. The binary number of adjustments can be read out to refer to the digital value 10. At the sampling moment of 100ns, $V_{out}$ is constant at the analog output voltage of 1.271V, which is converted to the mantissa code 01001 by a single slope. The resultant currents of different sizes show different slopes of the integral output curve $V_O$ and different numbers of adjustments before 100ns. The simulated Vout and digital output 1001001 are within credible error of their theoretical values of 1.28mV and 1001001. The system can realize the preset FP8 floating-point calculation function.

The linearity analysis of FP-DAC is shown in Fig. \ref{fig. 5}(b). We take $20uS$, $18uS$, $15uS$, and $12uS$ as RRAM conductance examples to obtain the device cell current with full coverage of the input pattern (0000000 to 1111111). Since FP8 (E2M5) is applied, the results are divided into 4 groups (00, 01, 10, and 11 for exponent). The evaluation results prove the correctness of cell current with different input patterns and weight conductance, showing good computing linearity of multiplication and MAC.

\begin{figure}[b]
\centerline{\includegraphics[scale=0.32]{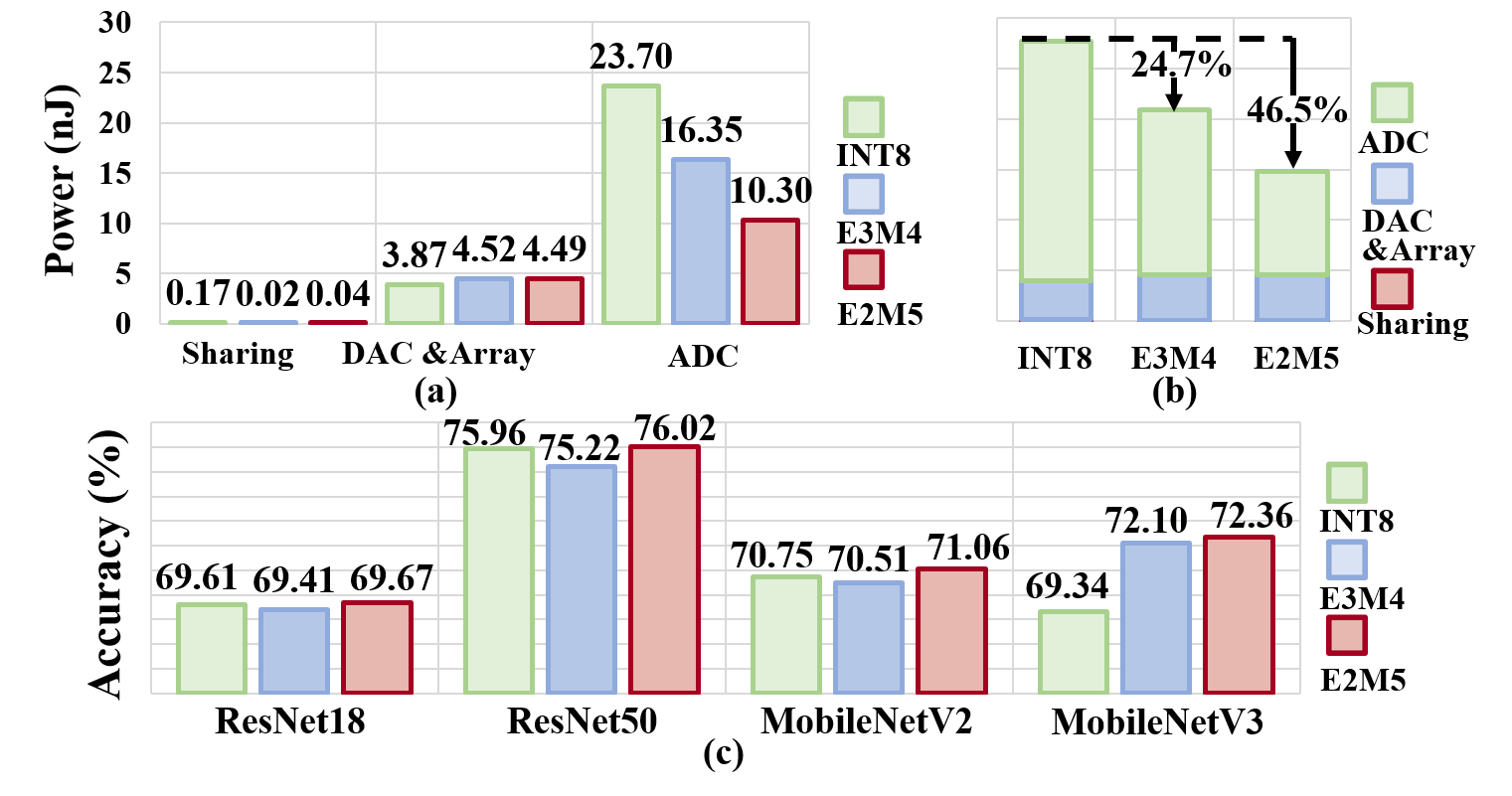}}
\caption{Comparison among INT8, FP8(E3M4) and FP8(E2M5) in (a) module power breakdown, (b) Total power, and (c) Top-1 accuracy in ImageNet}
\label{fig. 6}
\end{figure}

\subsection{Circuit Performance Analysis}

\begin{table*}[t]
\centering
\caption{CIM Macro Comparison}
\label{tab: core-level comparison} 
\renewcommand\arraystretch{1.2}
\scalebox{0.92}{
\begin{tabular}{|c|cc|c|c|c|c|c|}
\hline
                                      & \multicolumn{2}{c|}{\textbf{AFPR-CIM}}                       & Nature'22 \cite{wan2022compute}  & TCASI'20 \cite{zhang2020robust}   & ISSCC'22 \cite{tu202228nm}   & VLSI'21 \cite{lee202113}    & ISSCC'21 \cite{park20219}           \\ \hline
Architecture                          & \multicolumn{2}{c|}{\textbf{Analog-CIM}}                     & Analog-CIM & Analog-CIM & Digital-CIM & Digital-CIM & Digital Accelerator \\ \hline
Memory                                & \multicolumn{2}{c|}{\textbf{RRAM}}                           & RRAM       & RRAM       & SRAM        & SRAM        & -                   \\ \hline
Size                                  & \multicolumn{2}{c|}{\textbf{576*256}}                        & 256*256    & 256*256    & 128KB       & 160KB       & 293KB               \\ \hline
Technology(nm)                        & \multicolumn{2}{c|}{\textbf{65}}                             & 130        & 45         & 28          & 28          & 40                  \\ \hline
Supply Voltage (V)                    & \multicolumn{2}{c|}{\textbf{1.2-2.5}}                        & 1.8        & 1.1        & 0.6-1.0     & 0.76-1.1    & 0.75-1.1            \\ \hline
ADC                                   & \multicolumn{2}{c|}{\textbf{FP-ADC}}                          & Neuron     & SAR        & -           & -           & -                   \\ \hline
Activation Precision                  & \multicolumn{1}{c|}{\textbf{FP8(E2M5)}} & \textbf{FP8(E3M4)} & INT8       & INT8       & FP32        & BF16        & FP8                 \\ \hline
Macro Computing Latency(us)           & \multicolumn{1}{c|}{\textbf{0.2}}       & \textbf{0.15}      & 10.7       & 1.08       & -           & -           & -                   \\ \hline
Throughput(GOPS or GFLOPS)            & \multicolumn{1}{c|}{\textbf{1474.56}}   & \textbf{1966.08}   & 274        & 121.4      & 140         & 119.4       & 567                 \\ \hline
Energy Efficiency(TOPS/W or TFLOPS/W) & \multicolumn{1}{c|}{\textbf{19.89}}     & \textbf{14.12}     & 7          & 0.61       & 3.7         & 1.43        & 4.81                \\ \hline
\end{tabular}}
\end{table*}

In order to show the performance of the dynamic range adaptive idea proposed in this paper more fairly, we designed a conventional INT single-slope integral ADC in the same process. And we also designed the E3M4 hardware following the same pattern to comparatively demonstrate why we chose the E2M5 as the bit combination. Fig. \ref{fig. 6}(a) shows the power breakdown for the E2M5, E3M4, and INT hardware, with the power consumption calculated for all arrays at the same time.

The conversion accuracy of the integral ADC is determined by the integration time. In order to increase the 2-bit while maintaining the original accuracy in INT-ADC, it is necessary to increase the original readout time by $2^2=4$ times based on the original readout time of 100ns. Then the whole conversion time is increased from 200ns to 500ns, resulting in a 2.5× waste in power consumption. By limiting the circuit design requirements to a small range, the power of ADC could be reduced by $56.4\%$. Although the E3M4 is shorter in time compared to the E2M5, the power consumption is still higher due to the exponential increase in integrating capacitance, which leads to an exponential increase in the driving load and current of the op-amp. Besides, the ability to convert FP data is the major advantage of the proposed ADC compared to conventional ADCs.
The power of the array can be considered as the load power consumption of the input DACs. As demonstrated in Fig. \ref{fig. 6}(a), FP-DAC achieves FP expression while minimizing the additional power consumption. E2M5 achieves a very significant power consumption reduction in the INT and E3M4 formats. Compared to the total power consumption of INT8, E2M5 reduces the hardware power by $46.5\%$.

\subsection{Macro Specification Evaluation}

In order to demonstrate the advantages of the FP-CIM Macro proposed in this paper, the sparsity is derived in network simulation and then brought into the circuit for performance-power simulation. The performance at the Macro level is compared with other state-of-the-art designs. The main comparison work contains the digital domain FP-CIM design, the traditional FP8 accelerator, and the analog INT8 CIM work. As shown in Table \ref{tab: core-level comparison}, the AFPR-CIM shows an advantage in performance. The results show that the proposed design achieves a high energy efficiency of 19.89 TOPS/W and a throughput of 1474.56 GFLOPS in FP8 (E2M5) format. The data is in high-density mode at $0\%$ sparsity, which is also chosen for comparing the other works. FP8 accelerator and digital-domain FP-CIM work\cite{tu202228nm}\cite{lee202113}\cite{park20219}, due to the need for a product pipe-stage and product alignment, results in power dissipation. AFPR-CIM avoids this additional loss due to the analog method and offers a 4.135× and 5.376× improvement in energy efficiency, respectively; As for the analog INT8 CIM works \cite{wan2022compute}\cite{zhang2020robust}, the fixed-range ADC and sequential inputs also limit power efficiency and parallelism. Thanks to the analog computation and dynamic range adaptive FP-ADC strategy, AFPR-CIM achieves a 5.382× improvement in throughput and a 2.841× improvement in power efficiency.

\subsection{Network performance}
To verify the accuracy advantage of the FP8 (E2M5) format over INT8 and FP8 (E3M4) format, we extracted the nonlinearities in circuits and performed the accuracy simulation on the macro model simulator.  Fig. \ref{fig. 6}(c) shows the accuracy improvement of three formats over FP32 in post-train quantization (PTQ) form. As shown in the figure, in both Resnet and MobileNet models, E2M5 achieves higher accuracy than INT8. This is mainly due to the fact that the FP8 format significantly alleviates the consumption of the quantization process compared to INT8. Furthermore, the relatively wide dynamic range of FP8 also contributes positively to the accuracy. In addition, E2M5 also achieves higher accuracy than E3M4 due to the Gaussian distribution of several models. To offer a more comprehensive analysis, for well-behaved networks such as ResNet and MobileNet with few outliers, the extra 1bit of the exponential dynamic range of  E3M4 is excessive, and the 1bit fewer mantissa bits also means less accuracy. E2M5 combines the advantages of the other two formats to achieve higher accuracy with better network efficiency.

\section{CONCLUSION}
This paper proposed an analog domain FP8 floating-point CIM scheme to implement FP8 computation with high energy efficiency and an adaptively dynamic range. The proposed FP-ADC achieves adaptive matching of the input dynamic range through automatic capacitive reconfiguration and charge sharing. To further adapt floating-point algorithms at the interface and address the energy efficiency disadvantage of previous FP8 hardware, we enable FP-ADC to convert fixed-point analog domain computation results to FP8 (E2M5) digital codes.  Moreover,  we designed the corresponding FP-DAC to provide an analog representation of the FP activation. Finally, we present the proposed AFPR-CIM architecture and network mapping. This architecture achieves an energy efficiency of 19.89 TFLOPS/W and a throughput of 1474.56 GFLOPS at a computing latency of 200ns. Compared with the traditional FP8 Accelerator, digital FP-CIM, and analog INT8 CIM, the proposed AFPR-CIM achieves 4.135×, 5.376×, and 2.841× energy efficiency enhancement. Therefore, this work exhibits significance for further research on FP8 format and FP-CIM systems.

\bibliographystyle{IEEEtran}
\bibliography{IEEEabrv,ref}

\vspace{12pt}

\end{document}